\documentclass[12pt]{article}
\RequirePackage{ifthen}

\newboolean{shortversion}
\setboolean{shortversion}{false}

\newboolean{slacpub}
\setboolean{slacpub}{true}

\newboolean{worldscientific}
\setboolean{worldscientific}{false}

\RequirePackage{andreas}
\RequirePackage{feynmf/feynmp}

\newcommand{\thetitleis}{Quantization of the Coulomb Chain in an External Focusing Field}

\newcommand{\theabstractis}{
With the appropriate choice of parameters and sufficient cooling, charged particles in a circular accelerator are believed to undergo a transition
to a highly-ordered crystalline state\cite{Wei:1993}. The simplest possible crystalline
configuration is a one-dimensional chain of particles. In this paper, 
we write down the quantized version of its dynamics. We show that in
a low-density limit, the dynamics is that of a theory of interacting
phonons. There is an infinite sequence of $n$-phonon interaction terms,
of which we write down the first orders, which involve phonon scattering and
decay processes. The quantum formulation developed here can serve
as a first step towards a quantum-mechanical treatment of the
system at finite temperatures.
}

\ifthenelse{\boolean{worldscientific}}{
\author{Andreas~C.~Kabel}
\address{Stanford Linear Accelerator Center, \\
  2575 Sand Hill Road, Menlo Park, CA~94025\\
E-Mail: andreas.kabel@slac.stanford.edu}
}
\ifthenelse{\boolean{worldscientific}}{\title{\uppercase{\thetitleis}}}{}

\begin{document}

\ifthenelse{\boolean{slacpub}}{
\thispagestyle{empty}
\renewcommand{\thefootnote}{\fnsymbol{footnote}}

%%%%% Substitute your Pub number, month and year in the following:
%%
\begin{flushright}
{\small
SLAC--PUB--8760\\
January 2001\\}
\end{flushright}

\vspace{.8cm}

%%%%% Title and Author Information:
%%
\begin{center}
{\bf\large   
\thetitleis\footnote{Work supported by
Department of Energy contract  DE--AC03--76SF00515.}}

\vspace{1cm}

Andreas C. Kabel\\
Stanford Linear Accelerator Center, Stanford University,
Stanford, CA  94309\\

\vfill

{\bf\large   Abstract}

\end{center}

\begin{quote}
\theabstractis
\end{quote}

\vfill

%%%%%%%%%%%%%%%
%% Choose"Presented at," "Contributed to" for conference papers
%% or "Submitted to" for journal papers
%%%%%%%%%%%%%%%
\begin{center} 
{\it 18th Advanced ICFA Beam Dynamics Workshop on Quantum Aspects of Beam Physics}
{\it Capri, Italy}
{\it October 15--20, 2000}
\end{center}
}{}

\ifthenelse{\boolean{worldscientific}}{
\maketitle
\abstracts{\theabstractis}
}{}

\newcommand{\Lagrangian}{\mathcal{L}}

\section{Higher-Order Dynamics of the Coulomb Chain}

We consider an ensemble of charged point particles forced into an
one-dimensional setup by an external focusing field.  In equilibrium,
the particles will be equidistant longitudinally. We treat the limit
of an infinite, but periodic, chain. The problem will be treated in the rest
frame of an orbiting particle, curvature and retardation effects will
be neglected. 

The kinetic, potential and Coulomb interaction Lagrangian are, respectively
\begin{equation}\label{eq:lagrangian}
  \begin{aligned}
  \Lagrangian_k &= \frac1{2m}\sum_\mu \vec{\dot x_\mu}\vecdot{ \vec{\dot x_\mu}} \\
   \Lagrangian_p &= -\frac{m\omega^2_{ext,x}}2\sum_\mu (x_\mu^1)^2 -\frac{m\omega^2_{ext,y}}2\sum_\mu (x_\mu^2)^2 \\  
   \Lagrangian_i &= \frac12 \sum_{\mu\neq \nu} \Lagrangian_{\mu\nu} = 
    \frac{e^2}{2}\sum_{\mu\neq \nu}\frac{1}{\sqrt{(\vec x_\mu -\vec x_\nu +(\mu-\nu)\vec\lambda)^2}}
   \satzz,
  \end{aligned}
\end{equation}
where we have introduced local coordinates around each particle's
equilibrium position. The sums run over all lattice sites. $\vec\lambda$ is
the lattice vector, we use a coordinate system where
$\vec \lambda=(0,0,\lambda).$ The particle has mass $m$, and the external
focusing strengths are given by $\omega^2_{ext,x},\omega^2_{ext,y} $ and are
assumed to be constant along the ring. We are using natural units
with $\hbar=c=1$.

We expand (\ref{eq:lagrangian}) in $x_\mu$, that is, we write
\begin{equation}
  \Lagrangian_{\mu\nu} = _{p=0}^{\infty}\sum_{i_1,\ldots,i_p=1}^3\sum_{\mu_1,\ldots,\mu_p=1}^\infty \frac1{p!}\sum\Lagrangian_{\mu\nu}^{(p),\mu_1\cdots\mu_p}x_{\mu_1}^{i_1}\cdots x_{\mu_p}^{i_p}
 \satzz.
\end{equation}

$\Lagrangian_i^{(0)}$ diverges, but is irrelevant here; $\Lagrangian_i^{(1)}=0$, as the co{\"o}rdinates are expanded around
their equilibrium. For the first interesting
orders, we get
\begin{equation}
  \begin{braced}
    \begin{aligned}
      \Lagrangian^{(2),\mu_1\mu_2}_{\mu\nu} &= \frac{\Delta^1\Delta^2}{\lambda^3\abs{\mu-\nu}^3}\left(3\delta^{i_13}\delta^{i_23}-\delta^{i_1i_2}\right) \\
      \Lagrangian^{(3),\mu_1\mu_2\mu_3}_{\mu\nu} &= \frac{\Delta^1\Delta^2\Delta^3(\mu-\nu)}{\lambda^3\abs{\mu-\nu}^5}
   \sum_{\Pi(i)}\delta^{i_13}\left(\frac32\delta^{i_2i_3}-\frac52\delta^{i_1i_2i_3}\right) \\
      \Lagrangian^{(4),\mu_1\cdots\mu_4}_{\mu\nu} &= \frac{\Delta^1\Delta^2\Delta^3\Delta^4}{\lambda^5\abs{\mu-\nu}^5}\sum_{\Pi(i)}\left(
        \frac38 \delta^{i_1i_2}\delta^{i_3i_4}- \frac{15}4\delta^{i_1i_2}\delta^{i_33}\delta^{i_43} \right. \\
        &\left.
        +\frac{35}8\delta^{i_13}\delta^{i_23}\delta^{i_33}\delta^{i_43}\right)
            \end{aligned}
  \end{braced}\satzz,
\end{equation}
where we used the shorthand notation $\Delta^i=(\delta_{\mu_im}-\delta_{\mu_in})$.

Doing the summation over $m,n$, we get
\begin{equation}
  \begin{braced}
\begin{aligned}
  \Lagrangian^{(2)}_{\mu_1\mu_2} &= \left(2\delta^{\mu^1\mu^2}\sum_{\pm k=1}^\infty \Phi^{(3)}_{k_0} - 2\Phi^{(3)}_{\mu_1\mu_2}\right)\delta^{i_1i_2}(3\delta^{i_13}-1)\\
  \Lagrangian^{(3)}_{\mu_1\mu_2\mu_3} &= \sum_{\Pi(\mu)}\left( - \delta^{\mu_1\mu_2}\Phi^{(4)}_{\mu_1\mu_3}\right)
  \sum_{\Pi(i)}\delta^{i_13}\left(\frac32\delta^{i_2i_3} -\frac52\delta^{i_1i_2i_3}\right) \\
  \Lagrangian^{(4)}_{\mu_1\cdots\mu_4} &= \sum_{\Pi(\mu)}\left(\frac1{12}\delta^{\mu_1\mu_2\mu_3\mu_4}\sum_{\pm k=1}^\infty \Phi^{(5)}_{k0} - \frac13\delta^{\mu_1\mu_2\mu_3} \Phi^{(5)}_{\mu_1\mu_4}\right. \\
&\left. +\frac14\delta^{\mu_1\mu_2}\delta^{\mu_3\mu_4}\Phi^{(5)}_{\mu_1\mu_4} \right)
\\ &\times \sum_{\Pi(i)}\delta^{i_1i_2}\delta^{i_3i_4}\left(
        \frac38  + \delta^{i_43}\left( - \frac{15}4 +\frac{35}8\delta^{i_13}\right)
        \right)
\end{aligned}
  \end{braced}\satzz,
\end{equation}
where $\Phi^{(n)}_{\mu\nu}=\left.\frac12{e^2}\lambda^{-n}(\sgn(\mu-\nu))^{n-1}\abs{\mu-\nu}^{-n}\right|_{\mu\neq\nu}$ and $\Phi^{(n)}_{\mu\mu}=0$ and $\Pi$ denotes all permutations of a set of indices.

The sums over $\Phi_{k0}$ give
\begin{equation}
  \sum_{k=1}^\infty \Phi^{(n)}_{k0} = \frac 12 e^2\lambda^{-n}\zeta(n)
\end{equation}
for odd $n$ and vanish for even $n$.
($\zeta(3)\approx 1.202, \zeta(5)\approx1.037$).

As the interaction is translationally invariant, we proceed by Fourier transformation:
\begin{equation}
  \begin{braced}
    \begin{aligned}
      x^m_\mu = \frac1{2\pi}\int_{-\pi}^{\pi}e^{-ik\mu}\xi^m(k) \\
      \xi^m(k) = \sum_{\mu=-\infty}^{\infty}e^{ik\mu}x^m_\mu
    \end{aligned}
  \end{braced}
\end{equation}

We write down the interaction Lagrangian in this basis.
For convenience, we introduce vertex functions
\begin{equation}
  F^{(p)}_{i_1\ldots i_2} =
 \frac1{(2\pi)^{p-1}}
  \int   
  \delta^{2\pi}\left(\sum_{i=1}^pk_i\right)
  \tilde\Phi^{(p+1)}(k_{i_1}+\cdots+k_{i_n})\td^pk\satzz.
\end{equation}
Note that momentum conservation is only up to integer multiples of $2\pi$. After
some Fourier gymnastics, we have
\begin{equation}\label{eq:diagrams}
  \begin{braced}
\begin{aligned}
  \Lagrangian^{(2)} &= 2\left(F^{(2)} - F^{(2)}_1\right)\delta^{i_1i_2}(3\delta^{i_13}-1)\\
  \Lagrangian^{(3)} &= \sum_{\Pi(k)}F^{(3)}_1  \sum_{\Pi(i)}\delta^{i_13}\left(\frac32\delta^{i_2i_3} -\frac52\delta^{i_1i_2i_3} \right) \\
  \Lagrangian^{(4)} &= \sum_{\Pi(k)}\left(\frac1{12}F^{(4)} -\frac13F^{(4)}_{123}  +\frac14 F^{(4)}_{12} \right)
\\ &\sum_{\Pi(i)}\delta^{i_1i_2}\delta^{i_3i_4}\left(
        \frac38  + \delta^{i_43}\left(\frac{35}8\delta^{i_13}- \frac{15}4\right)\right)
\end{aligned}
  \end{braced}\satzz.
\end{equation}

\section{Quantization}

The quadratic terms of the total Lagrangian describe an
ensemble of harmonic oscillators with co{\"o}rdinate
variables $\xi_i(k), \xi^*_i(k)=\xi_i(-k)$. We introduce momenta variables
$\pi_i(k), \pi^*_i(k)=\pi_i(-k)$ obeying the usual commutation relations.

Quantization is straightforwardly done by defining 
creation and annihilation operators $a_i(k),a_i^{+}(k)$ by
\begin{equation}
  \sqrt{2\Omega(k)}a_j(k) = \Omega(k)\xi_j(k)+i\pi_j(-k)\satzz,
\end{equation}
with oscillator frequencies $\Omega(k)$ defined below.
These oscillator eigenmodes describe phononic (particle displacement
waves) excitations of our system. 

We write the full Lagrangian (\ref{eq:diagrams}) in terms of the operators $a_i(k),a_i^{+}(k)$.
The momentum-independent terms in (\ref{eq:diagrams}) are disposed of by  absorbing them into the Fourier transform of the potential:
   $\tilde\Phi(k)\to\tilde\Phi(k)-\tilde\Phi(0)$.

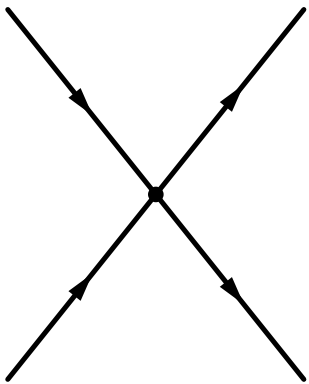
\begin{figure}\begin{center}
\begin{fmfgraph*}(140,140)
 \fmfpen{thick}
 \fmfleft{i1,i2}
 \fmfright{o1,o2}
 \fmf{fermion}{i1,v1}
 \fmf{fermion}{i2,v1}
 \fmf{fermion}{v1,o1}
 \fmf{fermion}{v1,o2}
 \fmflabel{$k_1$}{i1}
 \fmflabel{$k_2$}{i2}
 \fmflabel{$k_1+\Delta_k$}{o1}
 \fmflabel{$k_2-\Delta_k$}{o2}
 \fmfdot{v1}
 \fmflabel{$\frac14\tilde\Phi^{(5)}(\Delta k)$}{v1}
\end{fmfgraph*}
\end{center} 
\caption{Scattering diagram}\label{fig:scattering}
\end{figure}

Inspecting (\ref{eq:diagrams}), one notices that the
terms can be interpreted diagrammatically:
\begin{enumerate}
\item $F^{(2)}_1$ gives the one-particle propagator, \abk ie, it
 gives the dispersion relation $\Omega^2(k)$ for the phonons (Fig. \ref{fig:twoparticle})
\item $F^{(3)}_{1}$ describes a decay process: one incoming
   phonon decays into two outgoing ones (Fig. \ref{fig:decay1})
\item $F^{(4)}_{123}$ describes a decay process: one incoming
   phonon decays into three outgoing ones (Fig. \ref{fig:decay})
\item $F^{(4)}_{12}$ describes a scattering: two incoming
   phonons exchange momentum (Fig. \ref{fig:scattering})
\end{enumerate}

Note that our diagrams are in terms of the spatial co{\"o}rdinates $\xi, \xi^*$. If we
want to draw the diagrams in terms of phononic eigenmodes, we have to
use $\xi, \xi^*\propto a^+\pm a$ and draw all 8 possible three-point and 32 possible four-point
diagrams: Each leg in any of the diagrams can be flipped 
over to make an outgoing particle an ingoing one while changing the sign 
of its momentum.  

Also, we have to multiply each diagram by the polarization tensors,
\abk ie the totally symmetric $i_i$ dependent terms in (\ref{eq:diagrams}).
With an obvious notation for transverse and longitudinal polarizations,
these are given in Table \ref{tab:polarize}; contributions with index 
configurations not given in the table vanish.

\begin{table}[htbp]
  \begin{center}
    \begin{tabular}[t]{|c|c|}
      \hline Index Structure & Weight \\ \hline
      $(\perp,\perp)$ & $-1$ \\
      $(\parallel,\parallel)$ & $+2$ \\ \hline
      $(\parallel,\parallel,\parallel)$ & $-6$ \\
      $(\parallel,\perp,\perp)$ & $-2$ \\ \hline
      $(\perp,\perp,\perp,\perp)$ & $+9$ \\
      $(\perp,\perp,\perp',\perp')$ & $+3$ \\
      $(\perp,\perp,\parallel,\parallel)$ & $-12$ \\
      $(\parallel,\parallel,\parallel,\parallel)$ & $+24$ \\ \hline
    \end{tabular}
    \caption{Weight factors of different polarizations}
    \label{tab:polarize}
  \end{center}
\end{table}

\begin{fmffile}{Spectrum.feynman}

\begin{figure}
\begin{center}
\begin{fmfgraph*}(140,140)
 \fmfpen{thick}
 \fmfleft{i1}
 \fmfright{o1}
 \fmfbottom{s1}
 \fmf{fermion}{i1,v1}
 \fmf{fermion}{v1,o1}
 \fmfv{label=$k_1$}{i1}
 \fmfv{label=$k_1$}{o1}
 \fmfv{label.angle=-60,label=$\tilde\Phi^{(3)}(k_1)$}{v1}
\end{fmfgraph*}
\end{center}
\caption{Free two-point function}
\label{fig:twoparticle}
\end{figure}
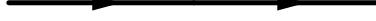

\begin{figure}\begin{center}
\begin{fmfgraph*}(140,140)
 \fmfpen{thick}
 \fmfleft{i1,i2}
 \fmfright{o1}
 \fmf{fermion}{i1,v1}
 \fmf{fermion}{i2,v1}
 \fmf{fermion}{v1,o1}
 \fmfdot{v1}
 \fmfv{label=$k_1$}{i1}
 \fmfv{label=$k_2$}{i2}
 \fmfv{label=$k_3=k_1+k_2$}{o1}
 \fmfv{label.angle=-60,label=$\tilde\Phi^{(4)}(k_3)$}{v1}
\end{fmfgraph*}
\end{center}
\caption{Decay diagram}
\label{fig:decay1}
\end{figure}

\begin{figure}\begin{center}
\begin{fmfgraph*}(140,140)
 \fmfpen{thick}
 \fmfleft{i1,i2,i3}
 \fmfright{o1}
 \fmf{fermion}{i1,v1}
 \fmf{fermion}{i2,v1}
 \fmf{fermion}{i3,v1}
 \fmf{fermion}{v1,o1}
 \fmfdot{v1}
 \fmfv{label=$k_1$}{i1}
 \fmfv{label=$k_2$}{i2}
 \fmfv{label=$k_3$}{i3}
 \fmfv{label=$k_4=k_1+k_2+k_3$}{o1}
 \fmfv{label.angle=-60,label=$-\frac13\tilde\Phi^{(5)}(k_4)$}{v1}
\end{fmfgraph*}
\end{center}
\caption{Decay diagram}
\label{fig:decay}
\end{figure}

\begin{figure}
\begin{center}
\begin{fmfgraph*}(140,140)
 \fmfpen{thick}
 \fmfleft{i1}
 \fmfright{o1}
 \fmfbottom{s1}
 \fmf{fermion}{i1,v1}
 \fmf{fermion}{v1,o1}
 \fmf{ghost}{s1,v1}
 \fmfdot{v1}
 \fmfv{label=$k_1$}{i1}
 \fmfv{label=$k_1+k_{ext}$}{o1}
 \fmfv{label=$k_{ext}$}{s1}
 \fmfv{label.angle=-60,label=$\tilde\Phi^{(3)}(k_1)+e^{i\frac{k\beta t}\lambda}K_{ext}(k_1)$}{v1}
\end{fmfgraph*}
\end{center}
\caption{Momentum insertion by the external lattice}
\label{fig:inject}
\end{figure}
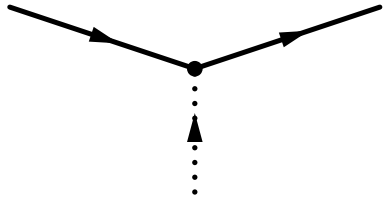

%\begin{figure}
%\begin{center}
%\begin{fmfgraph*}(80,80)
% \fmftop{i1,i2,i3,i4}
% \fmf{fermion}{i1,v1}
% \fmf{fermion}{i2,v1}
% \fmf{ghost}{v1,v2}
% \fmf{fermion}{i3,v2}
% \fmf{fermion}{i4,v2}
%\end{fmfgraph*}
%\end{center}
%\caption{}
%\label{}
%\end{figure}

%\begin{figure}
%\begin{center}
%\begin{fmfgraph*}(80,80)
% \fmftop{i1,i2,i3,i4}
% \fmf{fermion}{i1,v1}
% \fmf{fermion}{i2,v1}
% \fmf{ghost}{v1,v2}
% \fmf{fermion}{i3,v2}
% \fmf{fermion}{i4,v2}
%\end{fmfgraph*}
%\end{center}
%\caption{}
%\label{}
%\end{figure}

%\begin{figure}
%\begin{center}
%\begin{fmfgraph*}(80,80)
% \fmftop{i1,i2,i3}
% \fmfbottom{i4} 
% \fmf{fermion}{i1,v1}
% \fmf{fermion}{i2,v1}
% \fmf{ghost}{v1,v2}
% \fmf{fermion}{i3,v2}
% \fmf{fermion}{i4,v2}
%\end{fmfgraph*}
%\end{center}
%\caption{}
%\label{}
%\end{figure}

%\begin{figure}
%\begin{center}
%\begin{fmfgraph*}(80,80)
% \fmftop{i1,i2}
% \fmfbottom{i3,i4}
% \fmf{fermion}{i1,v1}
% \fmf{fermion}{i2,v1}
% \fmf{ghost}{v1,v2}
% \fmf{fermion}{i3,v2}
% \fmf{fermion}{i4,v2}
%\end{fmfgraph*}
%\end{center}
%\caption{}
%\label{}
%\end{figure}

%\begin{figure}
%\begin{center}
%\begin{fmfgraph*}(80,80)
% \fmftop{i1,i3}
% \fmfbottom{i2,i4}
% \fmf{fermion}{i1,v1}
% \fmf{fermion}{i2,v1}
% \fmf{ghost}{v1,v2}
% \fmf{fermion}{i3,v2}
% \fmf{fermion}{i4,v2}
%\end{fmfgraph*}
%\end{center}
%\caption{}
%\label{}
%\end{figure}

\end{fmffile}

Looking at the coefficient of the two-particle diagram, we can write
down the oscillator frequencies due to the internal
degrees of freedom:
\begin{equation}
  \Omega^2_k = \frac{e^2}{m\lambda^3}\sum_{\mu=1}^\infty \mu^{-3}(1-\cos(k\mu))\satzz.
\end{equation}

\begin{figure}[htbp]
  \begin{center}
    \includegraphics{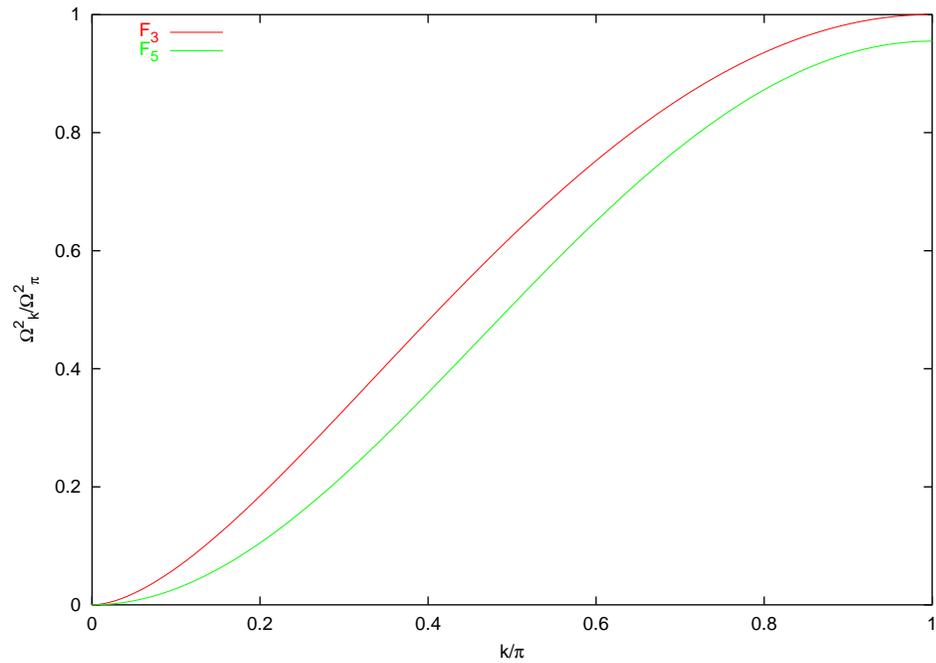}
    \caption{Spectrum and interaction strength of the infinite coulomb chain}
    \label{fig:spectrum}
  \end{center}
\end{figure}

The explicit form of this dispersion relation involves
generalized Zeta functions and is not too enlightening.
However, we can write down the energy of the $\pi$ mode, which is
easily seen to be the highest energy mode (see Fig. \ref{fig:spectrum}):
\begin{equation}\label{eq:freqmax}
  \Omega^2_\pi = \frac{e^2}{m\lambda^3}\sum_{\mu=1}^\infty 
  \mu^{-3}(1-(-1)^\mu) = 
  \frac{7e^2}{8m\lambda^3}\zeta(3)\satzz.
\end{equation}

One reads off that $\Omega^2_{\perp,k} = -\Omega^2_k $ and $\Omega^2_{\parallel ,k} = 2\Omega^2_k $, that is, in
the absence of external forces the transverse motion is unstable.
This has to be counteracted by an external focusing field with
a field gradient greater than $\Omega^2_\pi$. In real-world situations,
this field will be position-dependent, i.e., our Lagrangian ceases
to be diagonal in the Fourier basis. Instead, we have a convolution 
with the Fourier decomposition of the lattice focusing. Diagrammatically, this
means that the two-point functions can get injected momentum from 
the magnetic lattice (Fig. \ref{fig:inject}), the $K_{ext}(0)$ contribution
just being the average focusing strength.

Also, $\Omega(k)$ determines the validity of our quantization procedure.
Instead of quantizing the fermionic particles, we have quantized their
collective phononic excitations, which we obtained by expanding the
classical Lagrangian around the classical equilibrium (cf.
{\cite{Schulz:1993vs,Schulz:1992}). Obviously, the particles have to
  be localized even in the quantum-mechanical domain for this
  procedure to be valid.

We have seen the particles behave oscillator-like to lowest nontrivial
order. Thus, we can estimate their wave functions' longitudinal extension;
the ground state of an harmonic oscillator has
\begin{equation}
  \sigma^2=\frac1{m\omega}
\end{equation}
as its extension. For a point-particle expansion to be valid,
we have to require
\begin{equation}
  \frac1{m\Omega_{\pi}}\ll\lambda^2\satzz, 
\end{equation}
or, with (\ref{eq:freqmax})
\begin{equation}
  \label{eq:quantcond}
\frac{1}{e^2m} = r_{Bohr} \ll \frac{8\lambda}{7\zeta(3)} \approx \lambda\satzz,
\end{equation}
(which is the one-dimensional version of the $r_s\gg 1$ condition known from Wigner crystal theory\cite{Wigner:1934,Wigner:1938}),
so the quantization procedure is valid for low particle
densities. As the condition is expressed in the rest
frame (so $\lambda=\gamma\lambda_{Lab}$), the condition can easily be fulfilled in realistic
setups.

\section{Acknowledgments}
I wish to thank R.~Ruth and M.~Venturini for useful discussions.

%Writing 
%\begin{equation}
%  \pa k \Omega^2_k \propto \sum_{\mu=1}^\infty \sin(k\mu)\mu^{-2} = -\int_0^k\log\left(2\sin\frac t2\right)\td t\satzz,
%\end{equation}

%Now the integral on the right can be split in two parts with integrands of
%different signs:

%\begin{equation}
%\int_0^k\log\left(2\sin\frac t2\right)\td t = 
%\int_0^{\frac\pi3}\log\left(2\sin\frac t2\right)\td t
%+\int_{\frac\pi3}^k\log\left(2\sin\frac t2\right)\td t
%\end{equation}

%Estimating the second term by its maximum $(\pi-\pi/3)\log(2)$,

\ifthenelse{\boolean{worldscientific}}{\newcommand{\newblock}{}}{}

%\bibliography{/home/akabel/texlib/bibliography}

%
\end{document}